\documentclass[11pt]{article}

\usepackage[portrait, margin=1.25in]{geometry}
\usepackage{setspace}
\usepackage[utf8]{inputenc} 
\usepackage[T1]{fontenc}    
\usepackage{hyperref}       
\usepackage{url}            
\usepackage{booktabs}       
\usepackage{amsfonts}       
\usepackage{nicefrac}       
\usepackage{microtype}      

\usepackage{graphicx}
\usepackage[ruled,vlined]{algorithm2e}
\usepackage{indentfirst}
\usepackage{scalerel,stackengine}
\stackMath
\newcommand\reallywidehat[1]{%
\savestack{\tmpbox}{\stretchto{%
  \scaleto{%
    \scalerel*[\widthof{\ensuremath{#1}}]{\kern-.6pt\bigwedge\kern-.6pt}%
    {\rule[-\textheight/2]{1ex}{\textheight}}
  }{\textheight}%
}{0.5ex}}%
\stackon[1pt]{#1}{\tmpbox}%
}
\usepackage{appendix}
\usepackage{multicol}
\usepackage{amsmath}
{
      \newtheorem{assumption}{Assumption}
 }

\DeclareMathOperator*{\argmin}{arg\,min}

\usepackage{caption}
\usepackage{subcaption}
\usepackage{multirow}
\usepackage{ragged2e}
\usepackage{ dsfont }

\usepackage{lscape}

\usepackage{authblk}

\title{Maximizing Portfolio Predictability with Machine Learning\thanks{We thank Yongmiao Hong and Hayoue Yang for their helpful comments and suggestions.}}

\date{First draft: May 2020, \\ November 3, 2023}

\author[**]{Michael Pinelis}
\author[***]{David Ruppert}
\affil[**]{Department of Economics, Cornell University, \texttt{mdp93@cornell.edu}}
\affil[***]{Department of Statistics \& Data Science and School of Operations Research and Information Engineering, Cornell University, \texttt{dr24@cornell.edu}}

\begin{document}
\maketitle


\begin{abstract}
We construct the maximally predictable portfolio (MPP) of stocks using machine learning. Solving for the optimal constrained weights in the multi-asset MPP gives portfolios with a high monthly coefficient of determination, given the sample covariance matrix of predicted return errors from a machine learning model. Various models for the covariance matrix are tested. The MPPs of S\&P 500 index constituents with estimated returns from Elastic Net, Random Forest, and Support Vector Regression models can outperform or underperform the index depending on the time period. Portfolios that take advantage of the high predictability of the MPP's returns and employ a Kelly criterion style strategy consistently outperform the benchmark. 

\end{abstract}

\bigskip

\indent \indent \indent \ \ Convex Portfolio Optimization, Empirical Stock Pricing 

\text{}


\newpage 

\section{Introduction}
\noindent 

The maximally predictable portfolio (MPP) problem is originally proposed by Lo and MacKinlay (1997). Its weights maximize the fraction of variability in the portfolio return explained by its conditional expectation. The MPP can be easily obtained as the eigenvector with the smallest eigenvalue of the matrix which is the product of the inverse of the covariance matrix of forecasted return errors and the historical returns covariance matrix, as long as only the unity constraint for the weights is imposed on the optimization. However, when many realistic constraints are imposed as in practice, finding a globally optimal solution in a rigorous manner can be a hard task.

We revisit the problem and detail an algorithm by Gotoh and Fujisawa (2012) that efficiently computes the optimal portfolio weights for maximizing predictability under realistic constraints. With this algorithm, we compute the weights for MPPs of S\&P 500 index constituents. The key input is return forecasts from various machine learning models which identify the most predictable assets. More predictable assets may outperform or underperform the broader marker depending on the time period. Two types of portfolios formed from the MPP consistently outperform the index. First, we scale sorted returns top deciles by the MPP weights, giving stocks that have the greatest forecasted returns and have been the most predictable the most importance in the new portfolios. Second, we reward-risk time the MPP in a portfolio with the risk-free asset. 

An outline of the paper follows. Section \ref{sec:lit} reviews the literature. Section \ref{sec:headings} describes the portfolio optimization methodology, including the portfolio return predictability problem, constrained optimization, and model descriptions. Section \ref{sec:Empirical Results} demonstrates the results of using the MPP optimization algorithm and machine learning models, and Section \ref{sec:conclusion} concludes.

\section{Literature} \label{sec:lit}
Lo and MacKinlay (1997) first developed the concept of the MPP, maximizing the in-sample portfolio coefficient of determination. They use a linear factor model to estimate conditional return expectations for various assets and do not constrain the asset weights except for the nonnegative case. They attain high levels of portfolio predictability for asset groups consisting of market indexes, size-sorted portfolios, and sector-sorted portfolios. The approach, however, also leads to extreme asset weights. Our methodology differs in that non-linear models are used to obtain forecasted returns, the portfolio weights are constrained, and our investment universe is large-cap U.S. stocks.

There is a sizable stand of literature on the constrained optimization problem. Gotoh and Konno (2000) showed that a small scale problem with few assets can be solved within a practical amount of time by using classical Dinkelbach transformation and hyper-rectangular subdivision algorithm. Later, Yamamoto and Konno (2007) proposed a more efficient algorithm for solving the same problem by using 0-1 integer programming approach instead of a hyper-rectangular subdivision strategy. Yamamoto, Ishii, and Konno (2007) then use this algorithm to solve a maximal predictability portfolio (MPP) optimization problem with over a few hundred assets and up to 10 to 15 factors; they outperform the NIKKEI225 Index with standard factor models MPP over a limited time frame (about 100 months). Konno, Morita, and Yamamoto (2008) then reformulate the MPP problem with absolute deviation and find outperformance. Konno, Takaya, and Yamato (2010) apply a dynamic strategy for choosing the set of factors which fits best to the market data and see even better investment performance. Takay and Konno (2010) then apply turnover constraints and beat the index. A recent paper by Gotoh and Fujisawa (2012) propose an algorithm which can be implemented by simply solving a series of convex quadratic programs, and computational results show that it yields within a few seconds a (near) Karush–Kuhn–Tucker solution to instances in the previous papers which were solved via a global optimization method.

We apply the algorithm from Gotoh and Fujisawa (2012) but to the stocks in the S\&P 500 index. Rather than using standard linear models, we use Elastic Net, Random Forest, and Support Vector Regression.

There is also work that uses predictive errors from neural network models in portfolio optimization for Brazil stocks. Freitas et al. (2009) formulate a portfolio optimization model that minimizes the portfolio's overall prediction error. We consider this optimization objective and a variant as well for our investment universe.

To our knowledge, this is the first paper written on a machine learning approach to forming the MPP.

\section{Methodology}
\label{sec:headings}

We compute the weights that give the highest portfolio coefficient of determination with an algorithm that solves a series of convex optimization problems. The covariance matrix of predicted stock return errors from machine learning models is an input in the objective function. The weights of the stocks in the portfolio are estimated each month to maximize portfolio return predictability. We also form long-short and market-timing portfolios from the MPP weights. We use separate machine learning models to predict indiviudal stock volatilities for the implementation of the market-timing portfolios. The initial data the models are trained on are from 1962 to 1969. The strategies are then optimized on out-of-sample data from 1970 to April 1990 in a procedure called validation. Every 12 months, the training data shifts forward by one year and the models are refit. One set of models for each hyperparameter combination is kept in parallel. We select the combination of hyperparameters for each machine learning model that attains the highest overall predictive accuracy measured by out-of-sample $R^2$ over this validation period for all stocks. Then the models are tested on a holdout set from May 1990 to April 2020, data that provides a final estimate of the models’ performance after they have been validated, to prevent against backtest-overfitting (Bailey et al., 2015) \footnote{Holdout sets are never used to make decisions about which algorithms to use or for improving or tuning algorithms. Therefore, the performance on the holdout set is indicative of investment performance if an investor starts trading with the models and strategy today.}. This gives us a series of out-of-sample portfolio returns and corresponding performance metrics. Only one attempt on the holdout set is made. Algorithm 1 describes the general portfolio optimization approach.

\medskip
\begin{algorithm}[H] \label{alg:1}
\SetAlgoLined
 \For{each month $t=1$ to $T$}{
 \begin{enumerate}
 
 \item Update models with the data until the most \\ recent returns and predictors at time $t-1$

  \item{ \For{each stock $i=1$ to $n$}{
  \begin{enumerate}
  
  \item Forecast the return and volatility one month ahead

 \end{enumerate}
  }
  }
  
  \item  Recompute the covariance matrix of forecasted return errors and the covariance \\ matrix of observed returns using data only until time $t-1$
  
      \item Compute the optimal portfolio weights at time $t$ which give the highest \\ portfolio predictability
  \end{enumerate}
 }
 \caption{Portfolio Optimization Approach}
\end{algorithm}
\medskip


We conduct an extensive array of tests to evaluate the robustness of our results. A key result is that the typical investor can see substantial portfolio performance improvements in 1) weighting the top expected return deciles of stocks by their predictability rather than just equal weighting or by their expected returns and 2) timing the MPP instead of the market index. The predictability of the MPP is statistically and economically significant. The next section establishes the portfolio optimization problem.

\subsection{Maximizing Predictability}

This section defines the predictability of a portfolio and examines the case in Lo and MacKinlay (1997), followed by the specific models in our empirical implementation.

Consider the return vector of $n$ assets $\boldsymbol{R}_t = [r_{1t},r_{2t},...,r_{nt}]^\top $. The assets are those used to form the MPP. They can can be individual stocks or portfolios themselves, but in this paper they are the former. For the time being, assume the following:

\begin{assumption}
    $\boldsymbol{R}_t$ is a jointly stationary and ergodic stochastic process with finite expectation $E[\boldsymbol{R}_t] = \boldsymbol{\mu} = [\mu_{1},\mu_{2},...,\mu_{n}]^\top$ and autocovariance matrices $E[( \boldsymbol{R}_{t-k} - \boldsymbol{\mu})(\boldsymbol{R}_t- \boldsymbol{\mu})^\top] = \boldsymbol{\Sigma}_k,$ where $k \geq 0$ without loss of generality.
\end{assumption}

\noindent This assumption is made for notational simplicity, as stationarity allows to ignore time indices.

Let $\boldsymbol{Z}_t := \boldsymbol{R}_t - \boldsymbol{\mu}$ be a vector of de-meaned asset returns and $\boldsymbol{\hat{Z}}_t$ be the forecast of the de-meaned returns based on the information set at time $t-1$, $\mathcal{F}_{t-1}$. Now assume that 
\begin{equation}
    \boldsymbol{\hat{Z}}_t = E[\boldsymbol{Z}_t | \mathcal{F}_{t-1}],
\end{equation}
the conditional expectation of the asset returns with the information set at the previous time. $\boldsymbol{Z}_t$ can then be expressed as 
\begin{equation}
    \boldsymbol{Z}_t = \boldsymbol{\hat{Z}}_t + \boldsymbol{\epsilon}_t,
\end{equation}
where $E[\boldsymbol{\epsilon}_t|\mathcal{F}_{t-1}] = \boldsymbol{0} \in {\rm I\!R}^n$, $var[\boldsymbol{\epsilon}_t|\mathcal{F}_{t-1}] = \boldsymbol{\Gamma} \in {\rm I\!R}^{n \times n}$ and importantly for the following results $\boldsymbol{\hat{Z}}_t$ and $\boldsymbol{\epsilon}_t$ are assumed independent. In the information set $\mathcal{F}_{t-1}$ are observable leading economic variables like interest-rate spreads or stock-specific factors like momentum, fundamental company performance measures, liquidity, and volatility.  

Let $\boldsymbol{w}$ denote a linear combination of the assets in $\boldsymbol{Z}_t$. Then the predictability of this portfolio measured by the coefficient of determination is
\begin{equation}
    R^2(\boldsymbol{w}) = 1 - \frac{var[\boldsymbol{w}^\top\boldsymbol{\epsilon}_t]}{var[\boldsymbol{w}^\top\boldsymbol{Z}_t]} = \frac{var[\boldsymbol{w}^\top\boldsymbol{\hat{Z}}_t]}{var[\boldsymbol{w}^\top\boldsymbol{Z}_t]} = \frac{\boldsymbol{w}^\top \boldsymbol{\hat{\Sigma}}_{0} \boldsymbol{w}}{\boldsymbol{w}^\top \boldsymbol{\Sigma}_{0} \boldsymbol{w}},
\end{equation}
where $\boldsymbol{\hat{\Sigma}}_{0} = var[\boldsymbol{\hat{Z}}_t] = E[\boldsymbol{\hat{Z}}_t \boldsymbol{\hat{Z}}_t^\top ] \in {\rm I\!R}^{n \times n}$ and $\boldsymbol{\Sigma}_{0} = var[\boldsymbol{Z}_t] = E[\boldsymbol{Z}_t \boldsymbol{Z}_t^\top ] \in {\rm I\!R}^{n \times n}$. The $R^2(\boldsymbol{w})$ is the fraction of the variability in the portfolio return $\boldsymbol{w}^\top\boldsymbol{Z}_t$ explained by its conditional expectation, $\boldsymbol{w}^\top\boldsymbol{\hat{Z}}_t$.

Then maximizing predictability with respect to the relative asset weights $\boldsymbol{w}$,
\begin{equation}
   \max_{\boldsymbol{w}} \frac{\boldsymbol{w}^\top \boldsymbol{\hat{\Sigma}}_{0} \boldsymbol{w}}{\boldsymbol{w}^\top \boldsymbol{\Sigma}_{0} \boldsymbol{w}},
\end{equation}
it is straightforward to show that the solution to the optimization problem is the eigenvector $\boldsymbol{\lambda}^*$ of $ \boldsymbol{\Sigma}_{0}^{-1} \boldsymbol{\hat{\Sigma}}_{0} $ with the largest eignenvalue, which we denote $\lambda_{max}$. Then $\boldsymbol{w}^{*} = \boldsymbol{\lambda^*} / \boldsymbol{\lambda^*}^{\top}\boldsymbol{1} \in {\rm I\!R}^n$ is the normalized linear combination and gives the MPP. As an example, the MPP is derived for a multivariate linear model in the next subsection.

\subsubsection{Multivariate Linear Model}  \label{ARMA}

Starting with the model of returns as a linear function of $m$ predictors,
\begin{gather}
    \boldsymbol{Z}_t = \boldsymbol{\alpha} + \boldsymbol{B} \boldsymbol{X}_{t-1} + \boldsymbol{\epsilon}_t \\
     E[\boldsymbol{\epsilon}_t|\mathcal{F}_{t-1}] = \boldsymbol{0}, \ \ var[\boldsymbol{\epsilon}_t|\mathcal{F}_{t-1}] = \boldsymbol{\Gamma},
\end{gather}
where $\boldsymbol{\alpha} \in {\rm I\!R}^n$, $\boldsymbol{B} \in {\rm I\!R}^{n \times m}$ is the matrix of coefficients, $\boldsymbol{X}_{t-1} \in {\rm I\!R}^{m}$, and $\boldsymbol{\Gamma} \in {\rm I\!R}^{n \times n}$ is a positive definite covariance matrix, we can derive closed-form expressions for the covariances:
\begin{gather}
    var[\boldsymbol{\hat{Z}}_t] = \boldsymbol{B}var[\boldsymbol{X}_{t-1}]\boldsymbol{B}^{\top}  = \boldsymbol{\hat{\Sigma}}_0 \in {\rm I\!R}^{n \times n} \\
    var[\boldsymbol{Z}_t] = \boldsymbol{B}var[\boldsymbol{X}_{t-1}]\boldsymbol{B}^{\top}+ \boldsymbol{\Gamma} = \boldsymbol{\Sigma}_0 \in {\rm I\!R}^{n \times n}.
\end{gather}

\noindent The predictability maximization problem and the solution are the same.


The MPP has also been derived from the unconditional covariances. For applicability, the time-varying MPP can be constructed by replacing $\boldsymbol{\Sigma}_{0}$ and $\boldsymbol{\Gamma}$ with conditional counterparts. In the empirical implementation, each expression above can be estimated with historical data $ \boldsymbol{\Sigma}_{0}$ and $\boldsymbol{\Gamma} $ are estimated each month with rolling window sample data as sample covariance matrices. 

Equation 3 only holds if assuming the forecast errors are independent from the forecasts, which is often not realistic. Therefore, for the models we considered we will minimize the second term in Equation 3. This way, the optimization will directly depend on the covariances of differences between actual returns and their predictions in $\boldsymbol{\Gamma}$ which is described in the next section. 

\subsubsection{Maximizing Predictability with a General Model}

Consider the below specification with the model $f$ and matrix of observations $\boldsymbol{X} \in {\rm I\!R}^{n \times m} $:
\begin{gather}
    \boldsymbol{Z}_t = f_t( \boldsymbol{X}_t) + \boldsymbol{\epsilon}_t 
\end{gather}
In the implementation, $f_t$ denotes the model for the assets at time $t$ is updated every 12 months. Let each element of $\boldsymbol{R} \in {\rm I\!R}^{T \times n}  $ be the observed return $Z_{ti}$ and each element of $\boldsymbol{Q} \in {\rm I\!R}^{T \times n}$ be the forecasted return $\hat{Z}_{ti} = f_t( x_{1ti}, x_{2ti},   ... ,  x_{mti})$. Denote $\boldsymbol{E} = \boldsymbol{R} - \boldsymbol{Q} $. The $R^2(\boldsymbol{w})$ can be computed as 
\begin{equation} \label{priobj}
    1 - \frac{ \boldsymbol{w}^{\top} \boldsymbol{E}^{\top} \boldsymbol{E}   \boldsymbol{w} }{ \boldsymbol{w}^{\top} \boldsymbol{R}^{\top} \boldsymbol{R}   \boldsymbol{w}  }.
\end{equation} 


\noindent Minimizing the second term in (10) is the primary optimization objective.

We also consider similar objectives. Following Freitas et al. (2009) we also look at the following objective, minimizing the portfolio's overall forecast error without respect to the portfolio variance.

\begin{equation}
    \min \boldsymbol{w}^{\top} \boldsymbol{E}^{\top} \boldsymbol{E}   \boldsymbol{w}
\end{equation}

\noindent While the resulting portfolio's volatility may be larger, the portfolio's more extreme returns may be more easy to forecast correctly. 

\textbf{Theoretically the MPP can be important, but in practice it will only lead to predictable portfolios out-of-sample if the predictability of stocks does not change drastically in a short period of time. If the cross-sectional autocorrelation of prediction errors is high, that is, the recent past errors for stocks are relatively similar to next month's, then the out-of-sample portfolio predictability should be high. Alternatively, if the model errors for individual stocks are highly autocorrelated then the following simple approach to portfolio weighting should also lead to predictable performance:} take the mean of the past errors for each stock scaled to sum to one as the portfolio weights. The lookback windows are varied and discussed in the results section.

Lastly, we examine minimizing the ratio of the portfolio's forecast error and the expected return.

\begin{equation}
    \min  \frac{ \boldsymbol{w}^{\top} \boldsymbol{E}^{\top} \boldsymbol{E} \boldsymbol{w}   } { \boldsymbol{w}^{\top}  \boldsymbol{\mu} }
\end{equation}

\noindent To allocate to the best performing stocks at the right times, we will set the expected stock returns $\boldsymbol{\mu}$ as the predicted returns for the next month from the models.

While theoretically the MPP can be an important concept, in practice the weights are unrealistically extreme when unbounded. For this reason, the next section discusses the constrained optimization problem and details an efficient algorithm to find a near-optimal solution.

\subsection{Constrained Maximally Predictable Portfolio Optimization}

We consider constrained portfolio weights for real-life applicability.

\begin{assumption}
    The constraints on portfolio $\boldsymbol{w}$ are depicted by a bounded polyhedron of the form
\end{assumption}
\begin{equation}
    W = \{ \boldsymbol{w} \in {\rm I\!R}^n: \boldsymbol{1}^{\top}_n \boldsymbol{w} = 1, \boldsymbol{0} \leq \boldsymbol{w} \leq \boldsymbol{\bar{w}}, \boldsymbol{\mu}^{\top} \boldsymbol{w} \geq \rho  , A \boldsymbol{w} \leq \boldsymbol{b}  \}
\end{equation}
where $\boldsymbol{1}^{\top}_n := (1,...,1)^{\top} \in {\rm I\!R}^n $, $\boldsymbol{\bar{w}} \in {\rm I\!R}^n  $ is an upper bound vector, $\boldsymbol{\mu}^{\top}:= (\mu_1,...,\mu_n)^{\top} $ is a vector of estimated mean return, $\rho$ is a constant representing the expected return an investor requires, $\boldsymbol{A} \in {\rm I\!R}^{m \times n} $ and $\boldsymbol{b} \in {\rm I\!R}^{m} $. Then the constrained MPP is then obtained by solving:

\begin{center}
$
\begin{aligned}
\min_{\boldsymbol{w}} \quad & \frac{ \boldsymbol{w}^{\top} \boldsymbol{E}^{\top} \boldsymbol{E}   \boldsymbol{w} }{ \boldsymbol{w}^{\top} \boldsymbol{R}^{\top} \boldsymbol{R}   \boldsymbol{w}  }  \\
\textrm{s.t.} \quad & \boldsymbol{w} \in W  \\
\end{aligned}
$
\end{center}
\medskip

\noindent The objective is the same as for the unconstrained case shown previously. The algorithm we use to obtain the MPP is developed by (Gotoh and Fujisawa, 2012). Their approach, termed normalized linearization algorithm (NLA), solves a finite sequence of convex quadratic programs to reach a near-optimal solution. 

Let $\eta := 1/\sqrt{\boldsymbol{w}^{\top} \boldsymbol{R}^{\top} \boldsymbol{R}   \boldsymbol{w}}$ where the denominator is greater than zero and $y := \eta x$. We can then rewrite the problem as

\begin{center}
$
\begin{aligned}
 \min_{ \boldsymbol{y} } \quad &  \boldsymbol{y}^{\top} \boldsymbol{E}^{\top} \boldsymbol{E} \boldsymbol{y}  \\
\textrm{s.t.} \quad & \boldsymbol{\mu}^{\top}  \boldsymbol{y} \geq \rho \boldsymbol{1}_n^{\top} \boldsymbol{y}, \  \boldsymbol{0} \leq \boldsymbol{y} \leq \boldsymbol{1}_n^{\top} \boldsymbol{y} \boldsymbol{\bar{w}},  \ (\boldsymbol{A} - \boldsymbol{y} \boldsymbol{1}_n^{\top} ) \boldsymbol{y} \leq \boldsymbol{0},  \ \boldsymbol{y}^{\top} \boldsymbol{R}^{\top}   \boldsymbol{R} \boldsymbol{y} = 1   \\
\end{aligned}
$
\end{center}

\noindent Because of the single quadratic equality constraint, $\boldsymbol{y}^{\top} \boldsymbol{R}^{\top}   \boldsymbol{R} \boldsymbol{y} = 1$, the formulation above is a nonconvex quadratic program and still difficult to exactly solve. Now denote $\boldsymbol{u} = \boldsymbol{R} \boldsymbol{y}$. A key idea of the algorithm is to approximate the unit sphere $\boldsymbol{u}^{\top} \boldsymbol{u} = 1$ by its tangent hyperplane, $(\boldsymbol{u}^{k-1})^{\top} \boldsymbol{u}^{k-1} = 1$, at a point $\boldsymbol{u}^{k-1}$ in each iteration. One iteration of NLA solves the following convex optimization problem for a given point $(\boldsymbol{y}^{k-1},\boldsymbol{u}^{k-1})$ satisfying $(\boldsymbol{u}^{k-1})^{\top} \boldsymbol{u}^{k-1} = 1$. 

\begin{center}
$
\begin{aligned}
QP(\boldsymbol{u}^{k-1}): \ \ \ \min_{ \boldsymbol{y} } \quad &  \boldsymbol{y}^{\top} \boldsymbol{E}^{\top} \boldsymbol{E} \boldsymbol{y}  \\
\textrm{s.t.} \quad & \boldsymbol{\mu}^{\top}  \boldsymbol{y} \geq \rho \boldsymbol{1}_n^{\top} \boldsymbol{y}, \  \boldsymbol{0} \leq \boldsymbol{y} \leq \boldsymbol{1}_n^{\top} \boldsymbol{y} \boldsymbol{\bar{w}},  \ (\boldsymbol{A} - \boldsymbol{y} \boldsymbol{1}_n^{\top} ) \boldsymbol{y} \leq \boldsymbol{0},  \ \boldsymbol{y}^{\top} \boldsymbol{R}^{\top}  \boldsymbol{u}_{k-1} = 1   \\
\end{aligned}
$
\end{center}
The full algorithm is given below.

\medskip
\begin{algorithm}[H] \label{alg:2}
\KwResult{The optimal portfolio weights, $\boldsymbol{w}^*$ }

\begin{enumerate}
    \item Let $\epsilon > 0$ be a tolerance for termination. Let $(\boldsymbol{y}^0, \boldsymbol{u}^0)$ be a solution satisfying $(\boldsymbol{u}^0)^{\top} \boldsymbol{u}^0 = 1$, and set $k \leftarrow 1$.

    \item Solve the convex quadratic program $QP(\boldsymbol{u}^{k-1})$, and let $(\boldsymbol{\bar{y}}^k, \boldsymbol{\bar{u}}^k) $ be the obtained solution, where $\bar{\boldsymbol{u}}^k = \boldsymbol{R} \bar{\boldsymbol{y}}^k $.
    
    \item Set $(\boldsymbol{y}^k, \boldsymbol{u}^k) \leftarrow (\bar{\boldsymbol{y}}^k, \bar{\boldsymbol{u}}^k) / \sqrt{ (\bar{\boldsymbol{u}}^k)^{\top} \bar{\boldsymbol{u}}^k }  $. \smallskip
    
    \SetAlgoNoLine
    \eIf{ $ \sqrt{ (\bar{\boldsymbol{u}}^k)^{\top} \bar{\boldsymbol{u}}^k } \leq 1 + \epsilon $ }{ \Indp $\boldsymbol{w}^{*} = \boldsymbol{y}^k / (\boldsymbol{y}^k)^{\top} \boldsymbol{1}  $ }{ \Indp Set $ k \leftarrow k + 1$ and go to Step 1. }

\end{enumerate}

 \caption{Normalized Linearization}
\end{algorithm}

\smallskip

The finite convergence of this algorithm is proven in Gotoh and Fujisawa (2012). CVXR (Anqi et al., 2017) in R was employed in solving the convex quadratic
programs. The parameters are set to $\rho = -0.1$, $ \boldsymbol{A}$ as the identity matrix, and $\boldsymbol{b}$ as the ones vector. We allow the portfolio expected return to go below zero with the choice of $\rho$ because our interest is to form the most predictable portfolio and earn high returns in other ways. $\boldsymbol{u}^0$ is set to $\boldsymbol{R} (1/\sqrt{n}) \boldsymbol{1} $ and $\epsilon$ to $10^{-3}$. Although other interesting cases are possible, the parameter we examine here is $\boldsymbol{\bar{w}}$. 

We keep the same constraints for the objectives in (11) and (12). (11) is a standard convex optimization
problem. For (12) we modify the NLA algorithm appropriately. The next sections discusses the models used to forecast stock returns and volatilities.

\subsection{Elastic Net}  \label{ARMA}

Starting with a standard linear model,
\begin{equation}
     y_t = \mu + \sum_{i=1}^{m} \beta_i x_{i,t-1} + \epsilon_t
\end{equation}
where $y_t$ can be either excess return stock returns $R_{ti}$ or the volatility $\sigma_{ti}$, we can consider various forms of regression regularization to deal with the high dimensionality of the predictor set. This gives alternate procedures to estimate the model coefficients from OLS. First we describe LASSO, penalized regression that is designed to prevent overfitting with shrinkage.
 
To fit a model, minimize the objective function
\begin{equation}
    \min_{\mu,\beta_1,...,\beta_m } \frac{1}{T}  \sum_{t=1}^{T}  \frac{1}{N_t}  \sum_{k=1}^{N_t} \left (y_t - \mu - \sum_{j=1}^{m}\beta_j x_{j,t-1} \right )^2 + \lambda \sum_{j=1}^{m} |\beta_j|,
\end{equation}
where $\lambda \geq 0$ is the shrinkage parameter on the $l_1$ penalty. $N_t$ is the number of stocks or observations in the training data for month $t$. A higher value of $\lambda$ places a higher penalty on the coefficients' absolute values, selectively shrinking them, and a high enough $\lambda$ can make coefficients zero. This produces a looser fit on the training data but less chance of over-fitting in terms of out-of-sample forecasts. Setting $\lambda = 0$ gives the same coefficients as OLS. To select the optimal value, validation is typically done by testing the performance for a range of values on an out-of-sample data set. The parameter value that gives the maximum predictive accuracy is then used in the model on a distinct out-of-sample set for which results are reported.

While the LASSO fitting method typically improves predictions relative to the OLS model, it can sometimes select one predictor arbitrarily from a group of correlated predictors. Zou and Hastie (2005) proposed Elastic Net, regression with both $l_1$ and $l_2$ loss, which adds a second parameter and makes variable selection more robust. The objective function is
\begin{equation}
    \arg_{\mu,\beta_1,...,\beta_m } \frac{1}{T}  \sum_{i=1}^{T} \frac{1}{N_t}  \sum_{k=1}^{N_t} (y_t - \mu - \sum_{j=1}^{m}\beta_j x_{j,t-1} )^2 + \lambda ( \alpha \sum_{j}^{m} |\beta_j| + \frac{1}{2} (1-\alpha) \sum_{j=1}^{m} \beta_j^2 ) .
\end{equation}
The parameter $0 \leq \alpha \leq 1$ controls the blending of the $l_1$ and $l_2$ loss. Using $\alpha > 0$ results in a stronger tendency to select groups of correlated predictors. The parameters $\alpha$ and $\lambda$ for the Elastic Net model are chosen with the sample from 1970 to April 1990 with cross validation as described in Section \ref{sec:headings}. The out-of-sample predictions for LASSO or Elastic Net are given by
\begin{equation}
     \hat{y}_{t+1} = \hat{\mu} + \sum_{i=1}^{m} \hat{\beta_i} x_{i,t} .
\end{equation}
The predictions, like for a standard OLS linear model, are a weighted sum of variables. The next subsection discusses the machine learning model Random Forest, which relies on recursive partitioning of the feature space to make predictions, and why it can perform better than linear models in our portfolio allocation problem. 

\subsection{Random Forest}  \label{Random Forest}
Random Forest is an ensemble machine learning algorithm developed by Breiman (2001). The prediction by a Random Forest model is the majority vote across all the individual decision tree learners (Hastie et al., 2017). The default tree bagging procedure draws $B$ different bootstrap samples of the training data and fits a separate classification tree to the $bth$ sample. The forecast is the average of the trees' individual forecasts. Trees for a bootstrap sample are usually deep and overfit, meaning each has low bias but is inefficiently variable. Averaging over the $B$ predictions reduces the variance and stabilizes the trees’ forecast performance. Algorithm 2 gives the procedure used to construct a Random Forest with the implementation by Liaw and Wiener (2002).

\medskip
\begin{algorithm}
\KwResult{The ensemble of trees \{$T_b\}^B$ }

 \For{$b=1$ to $B$}{
  \begin{enumerate}
  \item Draw a bootstrap sample $\mathbf{Z^*}$ of size $n$ from the training data.
  
  \item Grow a random-forest tree $T_b$ to the bootstrapped data by \\ recursively repeating the following steps for each terminal node \\ of the tree, until the minimum node size fraction $s_{min}$ or the maximum \\ number of terminal nodes $k_{max}$ are reached.
  \begin{enumerate}
    \item Select $m$ variables at random from the $p$ variables
    
    \item Pick the best variable/split-point among the $m$.
    
    \item Split the node into two child nodes.
    \end{enumerate}
 \end{enumerate}
  
}
 \caption{Random Forest}
\end{algorithm}
\medskip
\noindent The prediction at a new point, $\boldsymbol{x}_t$, is 
\begin{equation}
  y_{t+1} = \hat{h}(\boldsymbol{x}_t) = \frac{1}{B} \sum_{b=1}^B \hat{T}_b(\boldsymbol{x}_t) ,
\end{equation} the average of all the individual trees' predictions. 

Random forests give an improvement over bagging with a variation designed to reduce the correlation among trees grown from different bootstrap samples. If most of the bootstrap samples are similar, the trees trained on these sample sets will be highly correlated. The average estimators of similar decision trees do not perform much better than a single decision tree. If, for example, among the variables, last month's dividend yield is the dominant predictor of the return, then most of the bagged trees will have low-depth splits on the most recent yield, resulting in a large correlation among their predictions. Trees are de-correlated with a method known as "random subspace" or "attribute bagging," which considers only a random subset of $m$ predictors out of $p$ for splitting at each potential branch. In the example, attribute bagging will ensure early branches for some trees will split on predictors other than the most recent dividend yield. Since each tree is grown with different sets of predictors, the average correlation among trees further decreases and the variance reduction relative to standard bagging is larger (Gu et al.\, 2020)\footnote{Because this makes Random Forest a non-deterministic algorithm, we average the results for multiple different seeds.}. The number of variables randomly sampled as candidates at each split, $m$, the number of bootstrap samples, $B$, the minimum fraction of observations in the terminal nodes, $s_{min}$, and $k_{max}$ are the tuning parameters optimized with validation. A detailed algorithm for classification trees can be found in the Appendix.

The parameters $m$, $s_{min}$, $k_{max}$, and $B$ are tuned with the sample from 1970 to April 1990. To test against parameter over-fitting, the final values are kept on the holdout time period from May 1990 to April 2020, for which results are reported, and only one attempt is made on the period. 

\subsection{Support Vector Regression} \label{Support Vector Regression}

First identified by Cortes and Vapnik (1995), support vector machine finds an optimal hyperplane between two classes. SVMs are used to predict stock returns in papers like Huerta et al. (2013). Support vector regression is the analog for real-valued response variables (Harris et al., 1996).

\begin{equation}
    y_{t+1} = f(\boldsymbol{x}) = \sum_{t=1}^T \sum_{k=1}^{N_t} \alpha_i z_i K(\boldsymbol{x}, \boldsymbol{x_{t,k}}) - b
\end{equation}
where $\boldsymbol{x_{t,k}}$ is the vector of observations for time $t$ and stock $t$, $T$ is the number of months in the training data, $N_t$ is the number of stocks in the training data for month $t$, $z_i$ is the response variable value, $b$ is a constant that shifts the predictions, and $\alpha_i$ is a scalar between 0 and C. C is the first parameter and controls how closely the model is fit to the training data. $\alpha_i$ and $b$ are given by solving the convex optimization problem described in the appendix. Lastly, $K(\boldsymbol{x_t}, \boldsymbol{x_{t,k}})$ is the kernel function. It is the radial basis function here.
\begin{equation}
    K(\boldsymbol{x}, \boldsymbol{x_{t,k}}) = e^{-\gamma \lVert \boldsymbol{x} - \boldsymbol{x_{t,k}} \rVert^2 / M }
\end{equation}
The kernel maps the feature vectors $\boldsymbol{x}$ into infinite dimensional space and takes the pair-wise distance between them. $\gamma > 0$ is the second parameter tuned. Higher values increase the influence of a single training example. The values used for the return models for $C$ and $\gamma$ are 3 and 0.1, respectively. For the volatility models, $C = 1$ and $\gamma = 0.1$. The R interface by Meyer (2023) to the well-known libsvm implementation (Change and Lin, 2022) is used.

%




\subsection{MPP Strategies}

\subsubsection{Timing the MPP}

In this section, we propose timing the MPP according to its prevailing reward and risk. The return forecasts from the models discussed in the previous subsections are utilized to maximize predictability in the portfolio. Taking the weighted sum of the individual stock return forecasts give an estimate for the expected return of the MPP. As the MPP is (by definition) the most predictable portfolio, predicting its return should easier than that of the market. We also forecast its monthly volatility.

Pinelis and Ruppert (2020) allocate between the market index and the risk-free asset with expected excess return and volatility forecasts from machine learning models. We adopt the same framework to determine the optimal weight of the MPP in a portfolio with the risk-free asset. The optimal weight of the MPP is
\begin{equation} \label{eq:12}
    w_t^{MT} = \frac{E[R_t-R_t^f|\mathcal{F}_{t-1}]}{\bar{\gamma} \cdot var[R_t |\mathcal{F}_{t-1}] },
\end{equation}
where $R_t -R_t^f$ is the excess return and $\bar{\gamma}$ represents the investor's level of risk-aversion which is set as 4.

Multiple estimates of the the excess return are considered. We use the weighted combination of individual monthly stock return forecasts as the conditional expectation, $E[R_t-R_t^f|\mathcal{F}_{t-1}] = \sum_{i=1}^n w_{it} E[R_{it} - R^f_{it} | \mathcal{F}_{t-1} ] $, where $w_{it}$ is the MPP weight of asset $i$ and time $t$. We could also fit a separate model to forecast the excess returns of the MPP. The features can be the same as for the individual stock model but simply calculated on the portfolio-level. That topic is left for further research. In risk timing, the volatility estimate is commonly a form of the realized monthly daily return variances. As the weighted sum of individual stock return forecasts is used for the MPP return forecast, the monthly volatility forecast is given by $ \sqrt{ \boldsymbol{w}_t^{\top} \boldsymbol{V} \boldsymbol{w}_t }$ where $\boldsymbol{V}$ is diagonal (we do not try to forecast covariances between stock returns) and has the individual stock squared volatility forecasts\footnote{An estimate for the MPP's volatility could aslo be given by a separate fitted stock volatility model with the same features and next month's realized monthly return volatility as the target variable for the stock volatility models.}. Then $w_t^{MT}$ can be computed as in Equation 21 and the final weights of the assets in the portfolio are $\boldsymbol{\alpha}_{t} = w_t^{MT} \boldsymbol{w}_t$.


\subsubsection{Long-Short Portfolios}

Next, rather than allocating between the MPP and the risk-free asset, we form a set of portfolios to directly exploit the varying predictability of different stocks' machine learning return forecasts. At the end of each month, one-month-ahead out-of-sample stock return predictions for each model are calculated. We then sort stocks into deciles based on each model’s forecasts. We reconstitute portfolios each month using equal weights and the MPP's weights. Finally, we construct a zero-net-investment portfolio that buys the highest expected return stocks (decile 10) and sells the lowest (decile 1). Therefore, there will be two sets of long-short portfolios which are identical in the set of holdings and have different weights. Weighting by the MPP can be thought of as a Kelly strategy (Kelly, 1956) where the MPP weights are proportional to the probability of successful bet. \textbf{The stocks with the highest expected returns and 'probabilities' of forecasting the high expected return correctly and will receive the highest weights and, if the probabilities are accurate, the MPP-weighted long-short portfolio should outperform the equal-weighted over time. In other words, the weights reflect the confidence in the forecasts.}

\section{Empirical Results} \label{sec:Empirical Results}

\subsection{Data}

This paper uses monthly data from CRSP, Compustat, and Kenneth French's\footnote{\url{http://mba.tuck.dartmouth.edu/pages/faculty/ken.french/data_library.html}} website. Stock prices, volume, and shares outstanding are from the CRSP database. Fundamental company data is from Compustat. Kenneth French's website contains the historical risk-free rate of return. The CRSP/Compustat merged database gives stock returns and the set of stocks that are in the S\&P 500 index each month.

From the data, fifteen fundamental features are formed. The features used in all the stock return models are a subset of those used in Gu et. al (2020) and are among the ones determined as most important for forecasting well, specifically: one-month momentum, market capitalization, six-month momentum, twelve-month momentum, change in momentum, maximum return, return volatility, change in shares, sales-to-price, share turnover, volatility of share turnover, price-to-earnings, book-to-market, and operating margin. For the stock volatility models, realized monthly lagged volatility is added to the variables listed before.

\subsection{MPP Predictability}

Here we show the characteristics of the MPP and variant portfolios for the different models. We can see that predictability varies significantly over time but is overall high.

Next, we examine the predictability of the portfolios. Table 1 compares the average monthly coefficient of determination (given by Equation 10) for the MPPs for the different models under constraint cases and the average $R^2$ for the model fits to asset returns.

\begin{table}[!htbp] 
\captionsetup{labelfont=bf,font=normalsize}
\caption{\textbf{Coefficient of Determination}}
\justifying{\small{\noindent This table contains the average portfolio R-squared (\%) from May 1990 to April 2020 for OLS, Elastic Net, and Random Forest models with the asset weights capped at 0.1, 0.3, and 0.5.  }}

\medskip

\centering
\label{tab:1}
\begin{tabular}{llll}
     \hline & & \multicolumn{1}{c}{ $\bar{w}_i $} & \\
\hline  
        ~ & 0.1 & 0.3 & 0.5 \\ \hline
        OLS & 20.97 & 21.48 & 21.48 \\ 
        Elastic Net & 16.31 & 16.61 & 16.61 \\ 
        Random Forest & 14.11 & 14.36 & 14.36 \\ 
    \end{tabular}
\end{table}

The predictability for portfolios formed with the different models is similar, but OLS actually attains the highest coefficient of determination over the three different weight maxes. This could be explained by the OLS model having a bigger range in the magnitude of prediction errors. In that case, the MPP optimization has greater freedom to allocate weights to stocks with very small prediction errors. Since the weights are chosen to minimize past forecast errors, this measure is not out-of-sample. 

To assess the out-of-sample predictive performance for individual stock return forecasts, we calculate the out-of-sample $R^2$ as

\begin{equation}
    R^2_{oos} = 1 - \frac{\sum_{t\in \mathcal{T} } ( r_{ti}- \hat{r}_{ti}  )^2  }{\sum_{t\in \mathcal{T} } r_{ti}^2}
\end{equation}
where $\mathcal{T}$ denotes the set of points not used for model training and $r$ are monthly excess returns relative to the S\&P500 index. Table 2 reports the average out-of-sample $R^2$ for each model.

\begin{table}[!htbp] 
\captionsetup{labelfont=bf,font=normalsize}
\caption{\textbf{Out-of-Sample Individual Stock Returns and Volatility R-squared}}
\justifying{\small{\noindent This table contains the average individual stock prediction R-squared (\%) from May 1990 to April 2020 for OLS, Elastic Net, Random Forest models.}}

\medskip

\centering
\label{tab:2}
    \begin{tabular}{lll}
        ~ & Returns & Volatility \\ \hline
        OLS & 0.07 & 50.99 \\ 
        Elastic Net & 0.22 & 50.46 \\ 
        Random Forest & 0.19 & 31.29 \\ 
    \end{tabular}
\end{table}

The monthly $R_{oos}^2$ is 0.19\% for the Random Forest model and 0.22\% for the Elastic Net model. This is  close to the benchmark out-of-sample $R^2$ that Gu et al. (2018) attain with Random Forest and Elastic Net on an optimized and comprehensive set of predictors. Forecasting volatility is easier as expected.

We can also see whether these result depend on the time period. The figure below plots the coefficients as a function of the time (rolling window of 36 months).

\begin{figure}[h]
\captionsetup{labelfont=bf,font=small}
  \centering
  \includegraphics[scale=.45]{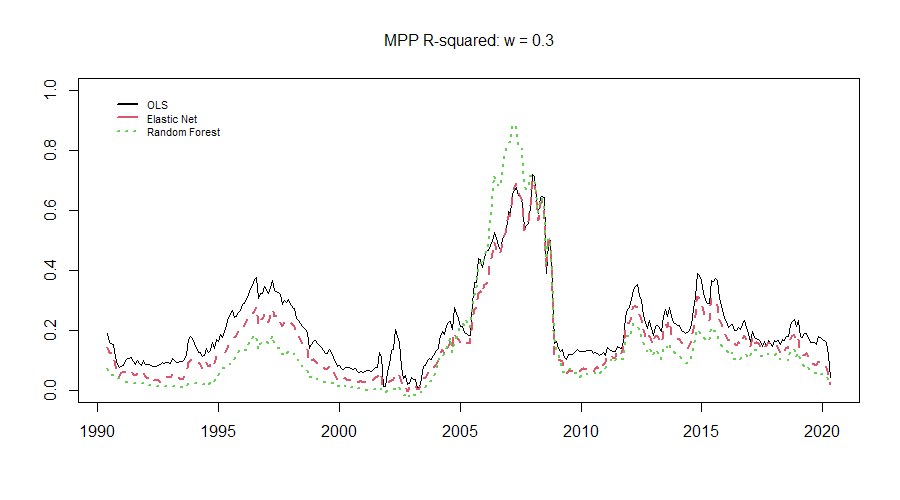}
  \caption{\textbf{Coefficient of Determination MPP Portfolios ($w_i = 0.3$)} This figure compares the predictability of the MPPs over time.}
  \label{fig:fig1}
\end{figure}

During stock market crises such as the Dot-Com bubble and the Great Recession expectantly the portfolios' predictabilities decrease. The predictability of the differnet models attain similiar levels over time.


We next look at how the predictability affects portfolio performance.


\subsection{Portfolio Performance}

This section discuss the out-of-sample investment performance for machine learning MPP and makes the relevant comparisons. We invest \$1 in May 1990 and plot the cumulative wealth for the portfolios, displayed in Figure 2. 

\begin{figure}[h]
\captionsetup{labelfont=bf,font=small}
  \centering
  \includegraphics[scale=.47]{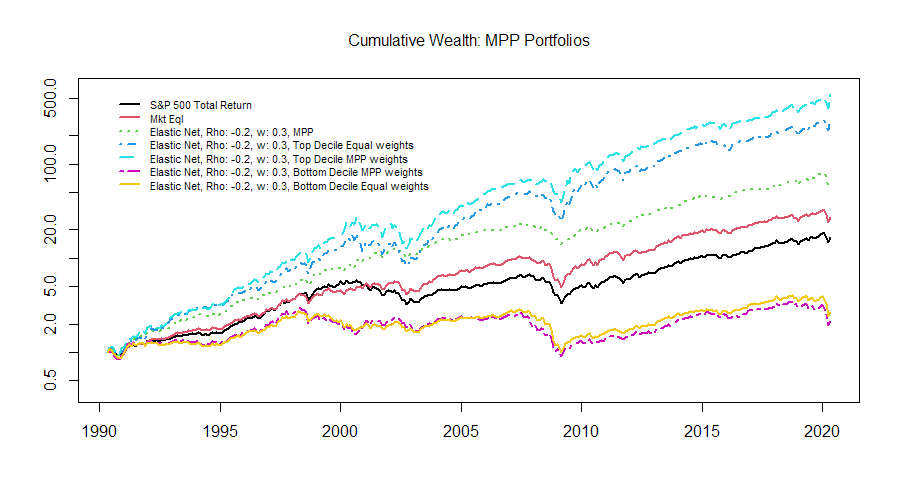}
  \caption{\textbf{Cumulative returns of MPP portfolios (Elastic Net, $w_i = 0.3$).} This figure plots the cumulative returns of the machine learning MPP portfolios from April 1990 to May 2020. The vertical axis is in log-scale.}
  \label{fig:fig2}
\end{figure}

The top decile of predicted returns weighted by MPP weights gives the best performance. The MPP gives a significant increase in investor's wealth, yet timing the machine learning MPP with its expected excess return and volatility (Figure 3) results in the best performance. The below figure shows the reward-risk timing for the Elastic Net MPP for $w_i = 0.3$. 

\begin{figure}[h]
\captionsetup{labelfont=bf,font=small}
  \centering
  \includegraphics[scale=.47]{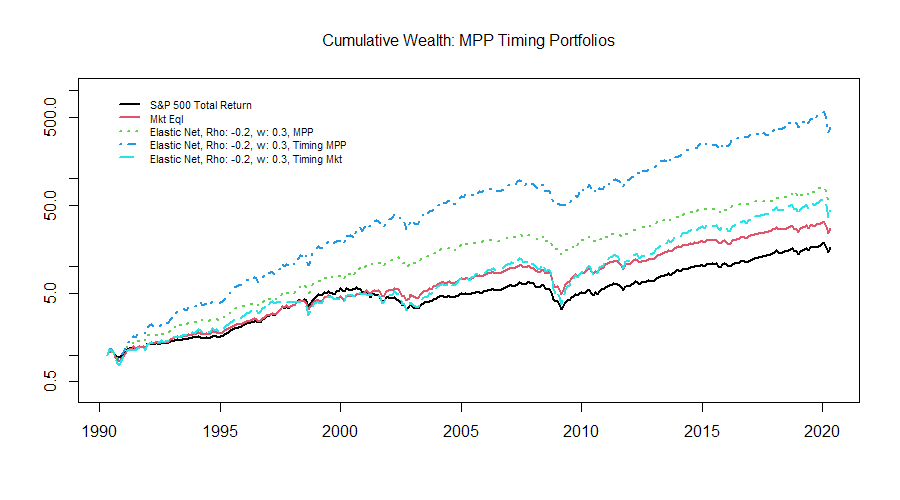}
  \caption{\textbf{Cumulative returns of Timing MPP portfolios (Elastic Net, $w_i = 0.3 $).} This figure plots the cumulative returns of the machine learning MPP portfolios from April 1990 to May 2020. The vertical axis is in log-scale.}
  \label{fig:fig2}
\end{figure}

Indeed, the MPP is more easily predictable and easier to reward-risk time.


Table 3 quantifies the risk-adjusted performance of the machine learning MPP, compared to the linear regression MPP and the buy-and-hold strategy for the two constraint cases.





\begin{table}[!htbp] 
\captionsetup{labelfont=bf,font=normalsize}
\caption{\textbf{Sharpe Ratios}}
\justifying{\small{\noindent  ($w_i = 0.3$). In this table are the out-of-sample annual returns, standard deviations, Sharpe ratios, and turnover for the holdout period from April 1990 to May 2020. Mkt denotes the buy-and-hold.}}

\medskip

\centering
\label{tab:3}
\resizebox{\textwidth}{!}{
\begin{tabular}{lllll}
        ~ & Annual Return (\%) & Standard Deviation (\%) & Sharpe & Turnover (\%) \\ \hline
        S\&P 500 & 10.46 & 14.6 & 0.54 & - \\ 
        Mkt Eql & 12.33 & 15.95 & 0.61 & - \\ 
        Mkt Tim & 15.36 & 15.95 & 0.61 & - \\ 
        OLS MPP & 13.81 & 14.66 & 0.76 & 53.21 \\ 
        OLS Min Err & 10.46 & 12.84 & 0.61 & 53.54 \\ 
        OLS Min Err/Ret & 8.96 & 14.46 & 0.44 & 98.67 \\ 
        OLS 1st Dec Eql & 20.77 & 22.5 & 0.81 & 76.54 \\ 
        OLS 10th Dec Eql & 5.45 & 17.34 & 0.16 & 86.5 \\ 
        OLS 1st Dec MPP & 20.99 & 22.63 & 0.81 & 103.15 \\ 
        OLS 10th Dec MPP & 4.12 & 18.25 & 0.08 & 138.26 \\ 
        OLS 1st Dec Err & 14.58 & 17.12 & 0.7 & 112.3 \\ 
        OLS 10th Dec Err & 6.1 & 15 & 0.23 & 128.57 \\ 
        OLS 1st Dec Err/Ret & 15.82 & 18.74 & 0.71 & 103.92 \\ 
        OLS 10th Dec Err/Ret & 3.23 & 17.11 & 0.04 & 124.3 \\ 
        OLS MPP Tim & 20.88 & 21.36 & 0.86 & 4.2 \\ 
        OLS Err Tim & 13.63 & 16.12 & 0.68 & 8.85 \\ 
        OLS Err/Ret Tim & 9.36 & 14.79 & 0.46 & 26.36 \\ \hline
        Elastic Net MPP & 15.06 & 15.23 & 0.82 & 55.7 \\ 
        Elastic Net Min Err & 10.41 & 12.85 & 0.61 & 54.11 \\ 
        Elastic Net Min Err/Ret & 10.62 & 14.88 & 0.54 & 87.11 \\ 
        Elastic Net 1st Dec Eql & 21.71 & 23.27 & 0.82 & 71.15 \\ 
        Elastic Net 10th Dec Eql & 4.62 & 16.15 & 0.12 & 70.54 \\ 
        Elastic Net 1st Dec MPP & 23.75 & 22.98 & 0.92 & 98.19 \\ 
        Elastic Net 10th Dec MPP & 4.12 & 17.49 & 0.09 & 125.4 \\ 
        Elastic Net 1st Dec Err & 17.28 & 18.07 & 0.81 & 106.83 \\ 
        Elastic Net 10th Dec Err & 5.06 & 13.84 & 0.18 & 107.75 \\ 
        Elastic Net 1st Dec Err/Ret & 18.3 & 20.19 & 0.78 & 103.05 \\ 
        Elastic Net 10th Dec Err/Ret & 4.7 & 15.45 & 0.14 & 112.03 \\ 
        Elastic Net MPP Tim & 22.44 & 22.13 & 0.9 & 2.96 \\ 
        Elastic Net Err Tim & 12.37 & 15.96 & 0.61 & 8.08 \\ 
        Elastic Net Err/Ret Tim & 13.4 & 17.29 & 0.62 & 15.86 \\  \hline
        Random Forest MPP & 14.28 & 15.87 & 0.74 & 61.4 \\ 
        Random Forest Min Err & 9.78 & 12.5 & 0.57 & 54.43 \\ 
        Random Forest Min Err/Ret & 9.56 & 14.79 & 0.47 & 68.87 \\ 
        Random Forest 1st Dec Eql & 18.92 & 26.61 & 0.61 & 50.99 \\ 
        Random Forest 10th Dec Eql & 8.02 & 15.85 & 0.34 & 64.26 \\ 
        Random Forest 1st Dec MPP & 19.31 & 27.22 & 0.61 & 89.96 \\ 
        Random Forest 10th Dec MPP & 8.35 & 15.82 & 0.36 & 107.95 \\ 
        Random Forest 1st Dec Err & 16.7 & 20.73 & 0.68 & 89.23 \\ 
        Random Forest 10th Dec Err & 6.05 & 13.33 & 0.26 & 100.42 \\ 
        Random Forest 1st Dec Err/Ret & 17.6 & 23.9 & 0.63 & 84.71 \\ 
        Random Forest 10th Dec Err/Ret & 8.25 & 15.79 & 0.36 & 83.22 \\ 
        Random Forest MPP Tim & 21.05 & 23.25 & 0.79 & 2.55 \\ 
        Random Forest Err Tim & 14.01 & 17.18 & 0.66 & 4.65 \\ 
        Random Forest Err/Ret Tim & 12.92 & 19.97 & 0.52 & 4.86 \\ 
    \end{tabular}
}
\end{table}

\noindent The machine learning MPP has higher annual returns than the market index. The appendix includes results for the other constraint cases. 

Overall, the results suggest that machine learning allows for more accurate realizations of which asset returns are most predictable and that the MPP can be used to reliably reward-risk time.


\section{Conclusion} \label{sec:conclusion}

We show that the maximally predictable portfolio (MPP) formed with machine learning is a significant source of return predictability. Using predicted stock returns from machine learning models to estimate the sample covariance of stock return errors leads to high portfolio R-squared values. While the MPP has higher predictability than the market index, when an investor times the machine learning MPP with its conditional excess return and volatility expectations, he or she can earn substantial improvements in risk-adjusted returns over not just the index but the portfolio which reward-risk times the index. Considering both the expected return for a stock and its predictability, specifically scaling the weight for highest expected return stocks by the confidence in their forecasts, leads to another level of portfolio outperformance.


\bibliographystyle{unsrt}  


\newpage

\appendix

\noindent {\LARGE \textbf{Appendix}}


\section{Regression Tree Algorithm}

Algorithm A1 details how to build a regression tree in a Random Forest and is a greedy algorithm (Breiman et al., 1984). We refer to the recursive version in (Murphy, 2012).
\bigskip

\setcounter{algocf}{0}
\renewcommand{\thealgocf}{A\arabic{algocf}}

\begin{algorithm}[H] 
\label{alg:3}
Initialize stump node, $N_1(0)$. $N_k(d)$ is the $k$th node at depth $d$. $S$ denotes the data, and $C$ is the set of unique labels. \\
\bigskip
function fitTree($N_k(d)$, $S$, $d$)
  \begin{enumerate}
  \item The prediction of the $N_k(d)$ node is the average value of its observations, $ \frac{1}{|N_k(d)|} \sum_{i\in N_k(d)} y_i $ 
  
  \item Define the cost function as the sum of squared differences from the mean: $cost(\{x_{i},y_{i}\}) = \sum_{i \in \{x_i,y_i\} } (y_i - \bar{y})^2 $, where $\bar{y} = \frac{1}{|\{x_i,y_i\}|} \sum_{i \in \{x_i,y_i\} } y_i $ \\ is the mean of the response variable in the specified set of data.
  
  \item Select the optimal split: \\ $(j^*, t^*) =  \argmin_{j \in \{1,..,m\} } \min_{t \in \mathcal{T}_j } (cost(\{ x_{i}, y_{i} :x_{ij} \leq t  \}) + cost(\{ x_{i}, y_{i} :x_{ij} > t \}) )$.
  
  $S_{left} = \{ x_{i}, y_{i} :x_{ij} \leq t  \}$, $S_{right} = \{ x_{i}, y_{i} :x_{ij} > t \}$.
  
  \item{\SetAlgoNoLine
     \eIf{notworthSplitting($d$, $cost$, $S_{left}$,$S_{right}$)}{
         \Indp  \ \ \ \ \  return $N_k(d)$
     }{
        \Indp{\Indp{
        Update the nodes: \\
         $N_1(d+1) = $ fitTree($N_k(d)$, $S_{left}$, $d+1$) \\
         $N_2(d+1) = $ fitTree($N_k(d)$, $S_{right}$, $d+1$) \\
         \ \ \ \ \   return $N_k(d)$ 
        }}
     }
  }      
 \end{enumerate}
 
\KwResult{The regression tree model $f(\vec{x}) = \sum_{m=1}^{D} w_m \mathds{1}\{ \vec{x} \in S_m \} $, where $w_m = \frac{1}{|S_m|} \sum_{i\in S_m} y_i  $ and $D$ is the number of regions }
 \caption{Regression Tree}
\end{algorithm}
\medskip
\noindent The function $notworthSplitting$($d$, $cost$, $S_{left}$,$S_{right}$) contains stopping heuristics to prevent overfitting. In our case, the function value is true if the fraction of examples in either $S_{left}$ or $S_{right}$ is less than $s_{min}$, the minimum fraction of observations in a node for a split determined by the user's parameter optimization, or if the number of terminal nodes $D$ is equal to $k_{max}$, the maximum number of terminal nodes. An important note is that the $S_{min}$ threshold is applied to the current node. For instance, a node can contain 5 observations out of 100 in the data even if $S_{min} = 0.9$, but any further splits from that node will not be made since $5/100 < 0.9$.

For the return models, the values we set for $s_{min}$, $k_{max}$, the number of trees, and the number of variables to select from at each split ($m$) are 0.95, 6, 500, and 5, respectively. For the volatility models, the values we set for $s_{min}$, $k_{max}$, the number of trees, and the number of variables to select from at each split ($m$) are 0.92, 8, 500, and 5, respectively.

\section{Support Vector Regression Convex Optimization}

The SVR model requires the solution of the following optimization problem:
\begin{align*}
& \min_{\boldsymbol{\alpha}, b, \boldsymbol{\mathcal{E}}} \frac{1}{2}  \sum_{i=1}^N \sum_{j=1}^N \alpha_i \alpha_j K(\boldsymbol{x}_i, \boldsymbol{x}_j) + C \sum_{i=1}^N \mathcal{E}_i  \\
& \textrm{s.t.}  \quad z_i \left( \sum_{j=1}^N \alpha_j K(\boldsymbol{x}_i, \boldsymbol{x}_j) + b \right) \geq 1 - \xi_i \quad \text{for } i = 1, 2, \ldots, N,  \\
& \ \ \ \  \quad \xi_i \geq 0 \quad \text{for } i = 1, 2, \ldots, N,
\end{align*}
where $\boldsymbol{\alpha}$ are the weights, $b$ is the bias, $N$ is the number of training examples, and $K$ is the Gaussian kernel function. $C$ is the cost hyperparameter and $\gamma$ is the radial basis function hyperparameter. For the return models, we set $C = 3$ and $\gamma = 0.1$ and for volatility models $C = 3$, $\gamma = 0.1$.

\section{Additional Figures \& Tables}

\begin{figure}[h]
\captionsetup{labelfont=bf,font=small}
  \centering
  \includegraphics[scale=.45]{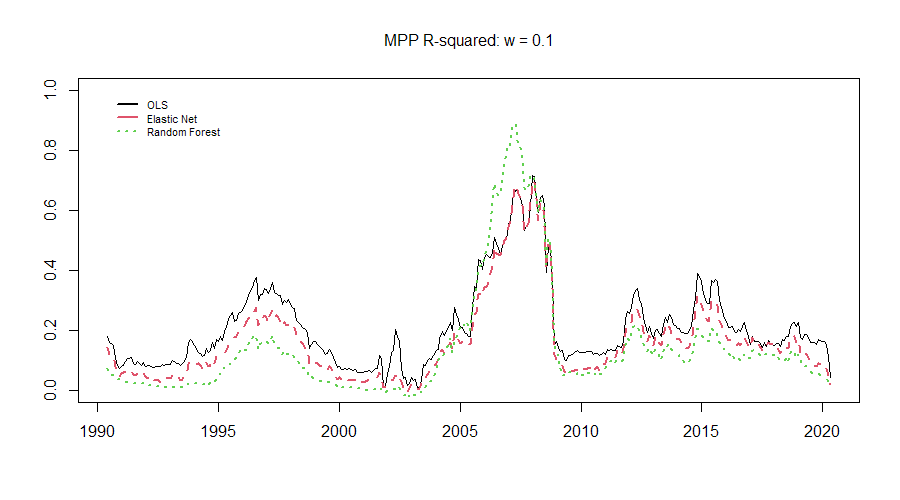}
  \caption{\textbf{Coefficient of Determination MPP Portfolios ($w_i = 0.1$)} This figure compares the predictability of the MPPs over time.}
  \label{fig:fig1}
\end{figure}

\begin{figure}[h]
\captionsetup{labelfont=bf,font=small}
  \centering
  \includegraphics[scale=.45]{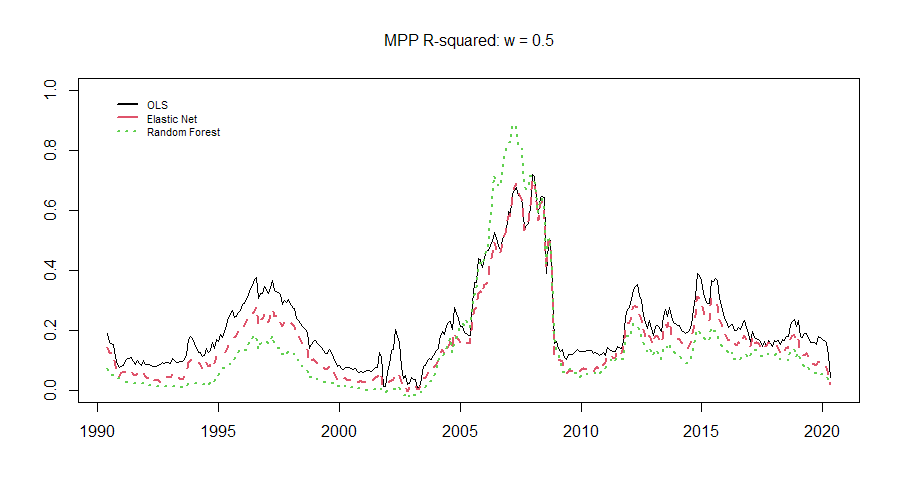}
  \caption{\textbf{Coefficient of Determination MPP Portfolios ($w_i = 0.5$)} This figure compares the predictability of the MPPs over time.}
  \label{fig:fig1}
\end{figure}

\begin{figure}[h]
\captionsetup{labelfont=bf,font=small}
  \centering
  \includegraphics[scale=.47]{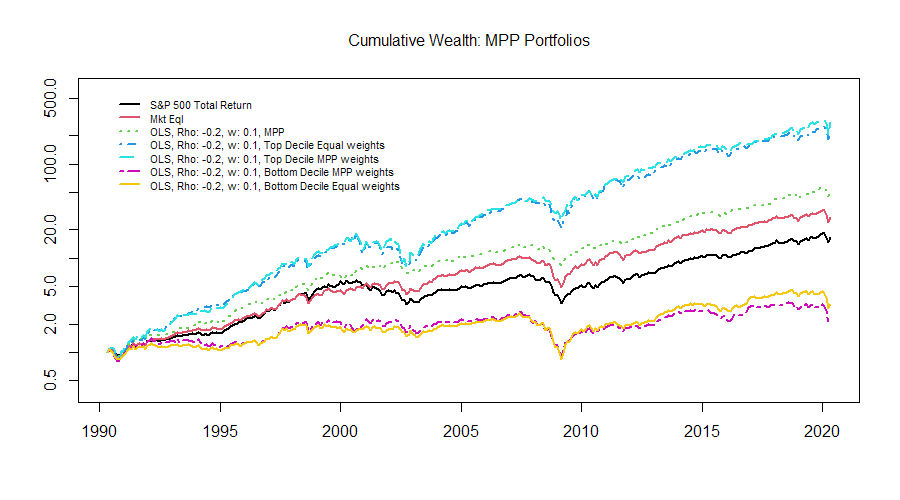}
  \caption{\textbf{Cumulative returns of MPP portfolios (OLS, $w_i = 0.1$).} This figure plots the cumulative returns of the machine learning MPP portfolios from April 1990 to May 2020. The vertical axis is in log-scale.}
  \label{fig:fig2}
\end{figure}

\begin{figure}[h]
\captionsetup{labelfont=bf,font=small}
  \centering
  \includegraphics[scale=.47]{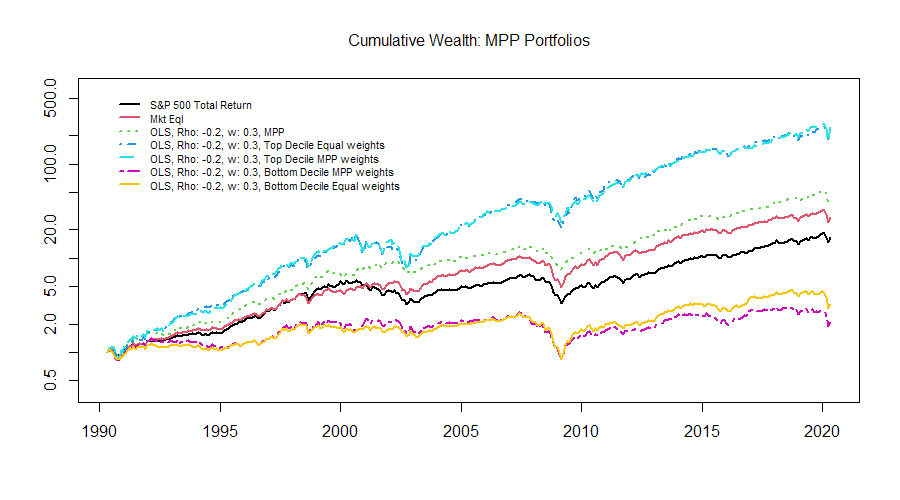}
  \caption{\textbf{Cumulative returns of MPP portfolios (OLS, $w_i = 0.3$).} This figure plots the cumulative returns of the machine learning MPP portfolios from April 1990 to May 2020. The vertical axis is in log-scale.}
  \label{fig:fig2}
\end{figure}

\begin{figure}[h]
\captionsetup{labelfont=bf,font=small}
  \centering
  \includegraphics[scale=.47]{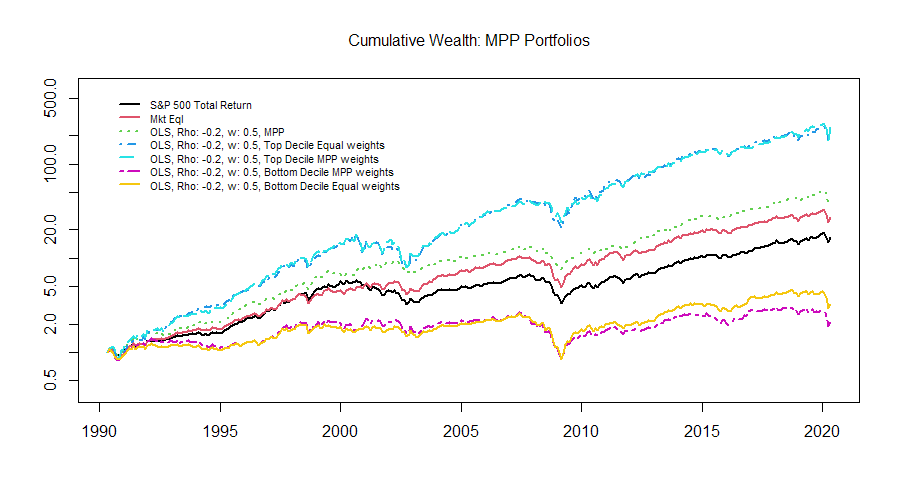}
  \caption{\textbf{Cumulative returns of MPP portfolios (OLS, $w_i = 0.5$).} This figure plots the cumulative returns of the machine learning MPP portfolios from April 1990 to May 2020. The vertical axis is in log-scale.}
  \label{fig:fig2}
\end{figure}

\begin{figure}[h]
\captionsetup{labelfont=bf,font=small}
  \centering
  \includegraphics[scale=.47]{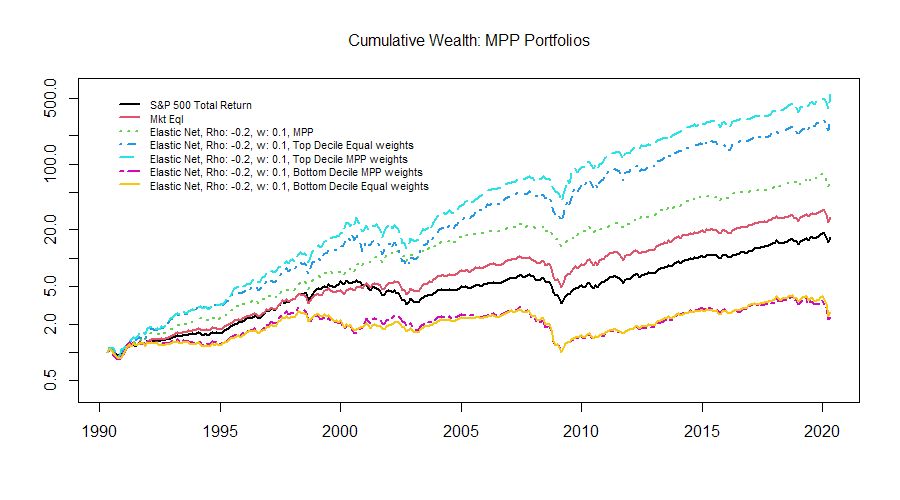}
  \caption{\textbf{Cumulative returns of MPP portfolios (Elastic Net, $w_i = 0.1$).} This figure plots the cumulative returns of the machine learning MPP portfolios from April 1990 to May 2020. The vertical axis is in log-scale.}
  \label{fig:fig2}
\end{figure}

\begin{figure}[h]
\captionsetup{labelfont=bf,font=small}
  \centering
  \includegraphics[scale=.47]{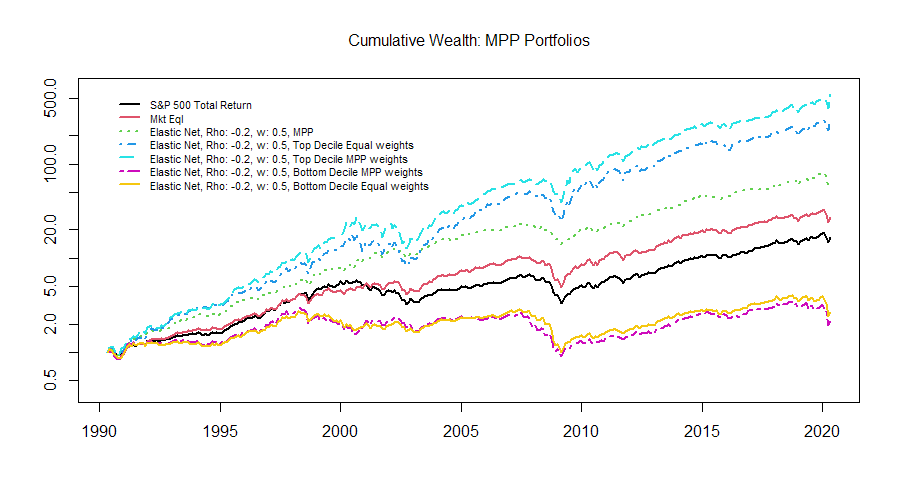}
  \caption{\textbf{Cumulative returns of MPP portfolios (Elastic Net, $w_i = 0.5$).} This figure plots the cumulative returns of the machine learning MPP portfolios from April 1990 to May 2020. The vertical axis is in log-scale.}
  \label{fig:fig2}
\end{figure}

\begin{figure}[h]
\captionsetup{labelfont=bf,font=small}
  \centering
  \includegraphics[scale=.47]{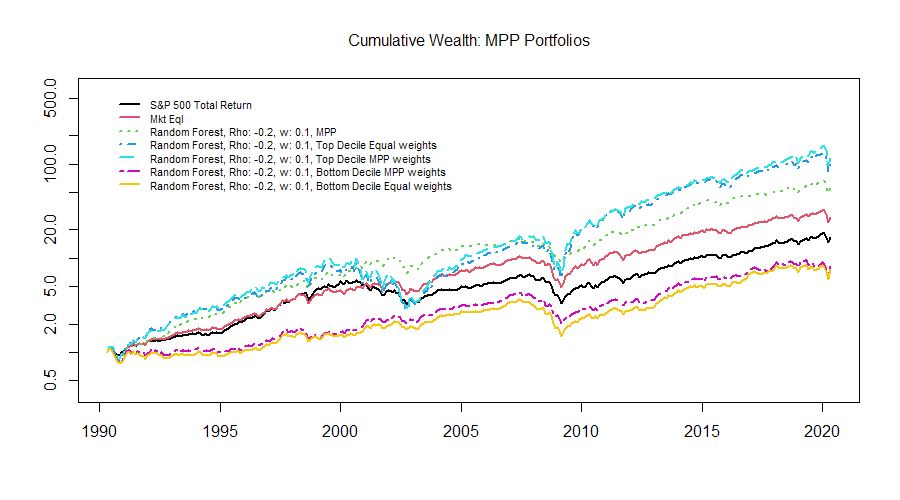}
  \caption{\textbf{Cumulative returns of MPP portfolios (Random Forest, $w_i = 0.1$).} This figure plots the cumulative returns of the machine learning MPP portfolios from April 1990 to May 2020. The vertical axis is in log-scale.}
  \label{fig:fig2}
\end{figure}

\begin{figure}[h]
\captionsetup{labelfont=bf,font=small}
  \centering
  \includegraphics[scale=.47]{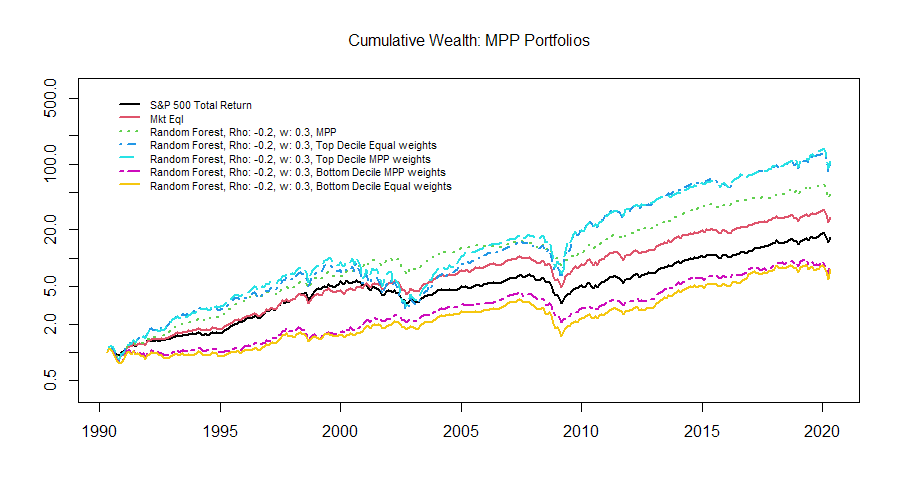}
  \caption{\textbf{Cumulative returns of MPP portfolios (Random Forest, $w_i = 0.3$).} This figure plots the cumulative returns of the machine learning MPP portfolios from April 1990 to May 2020. The vertical axis is in log-scale.}
  \label{fig:fig2}
\end{figure}

\begin{figure}[h]
\captionsetup{labelfont=bf,font=small}
  \centering
  \includegraphics[scale=.47]{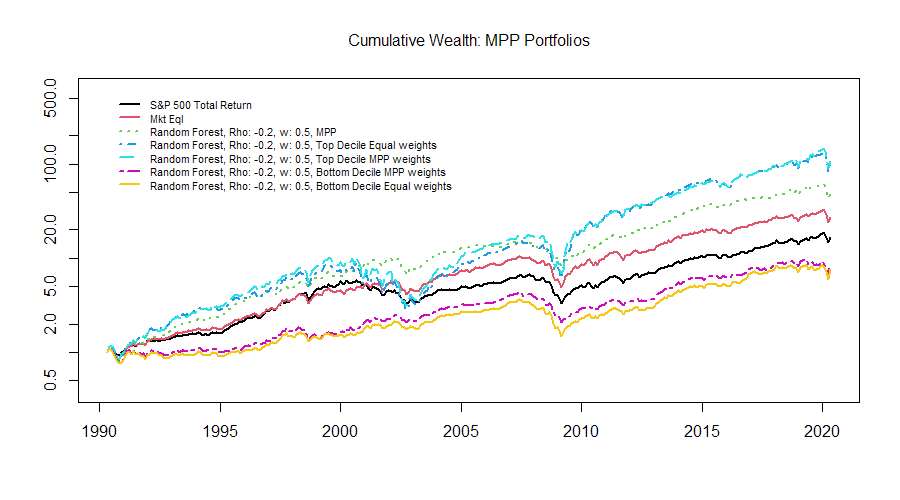}
  \caption{\textbf{Cumulative returns of MPP portfolios (Random Forest, $w_i = 0.5$).} This figure plots the cumulative returns of the machine learning MPP portfolios from April 1990 to May 2020. The vertical axis is in log-scale.}
  \label{fig:fig2}
\end{figure}

\begin{figure}[h]
\captionsetup{labelfont=bf,font=small}
  \centering
  \includegraphics[scale=.47]{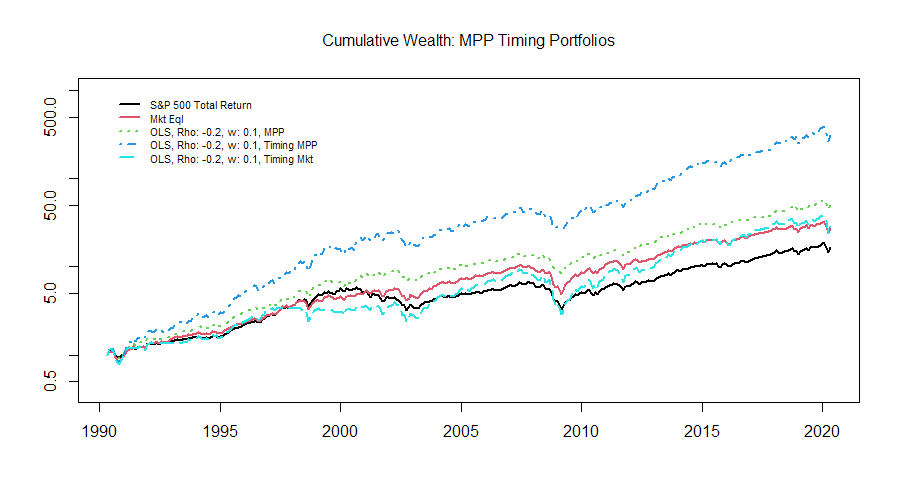}
  \caption{\textbf{Cumulative returns of Timing MPP portfolios (OLS, $w_i = 0.1 $).} This figure plots the cumulative returns of the machine learning MPP portfolios from April 1990 to May 2020. The vertical axis is in log-scale.}
  \label{fig:fig2}
\end{figure}

\begin{figure}[h]
\captionsetup{labelfont=bf,font=small}
  \centering
  \includegraphics[scale=.47]{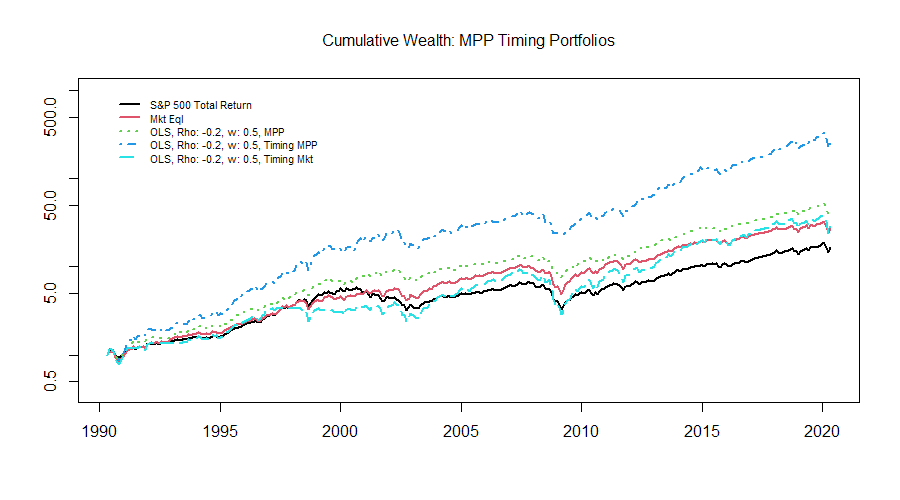}
  \caption{\textbf{Cumulative returns of Timing MPP portfolios (OLS, $w_i = 0.3 $).} This figure plots the cumulative returns of the machine learning MPP portfolios from April 1990 to May 2020. The vertical axis is in log-scale.}
  \label{fig:fig2}
\end{figure}

\begin{figure}[h]
\captionsetup{labelfont=bf,font=small}
  \centering
  \includegraphics[scale=.47]{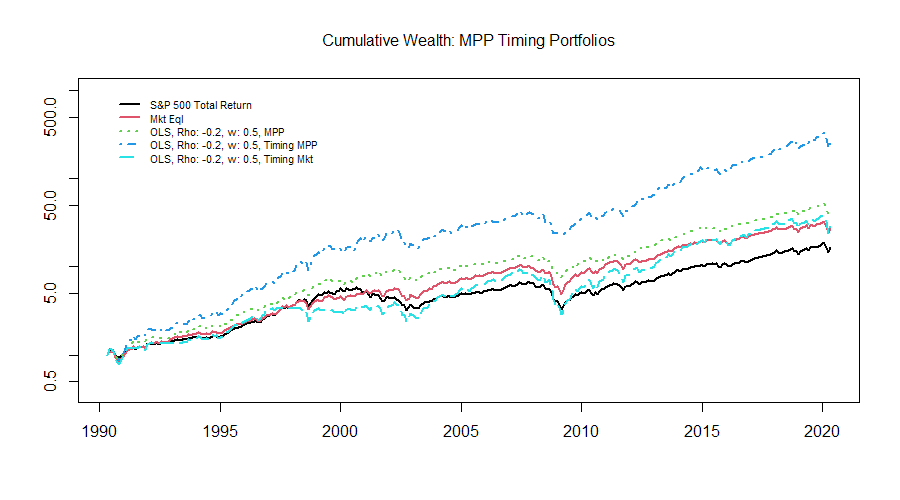}
  \caption{\textbf{Cumulative returns of Timing MPP portfolios (OLS, $w_i = 0.5 $).} This figure plots the cumulative returns of the machine learning MPP portfolios from April 1990 to May 2020. The vertical axis is in log-scale.}
  \label{fig:fig2}
\end{figure}

\begin{figure}[h]
\captionsetup{labelfont=bf,font=small}
  \centering
  \includegraphics[scale=.47]{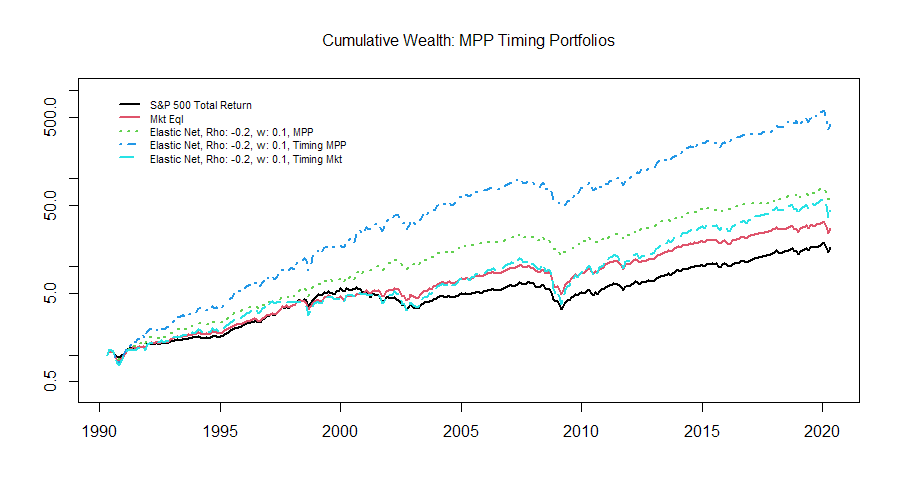}
  \caption{\textbf{Cumulative returns of Timing MPP portfolios (Elastic Net, $w_i = 0.1 $).} This figure plots the cumulative returns of the machine learning MPP portfolios from April 1990 to May 2020. The vertical axis is in log-scale.}
  \label{fig:fig2}
\end{figure}

\begin{figure}[h]
\captionsetup{labelfont=bf,font=small}
  \centering
  \includegraphics[scale=.47]{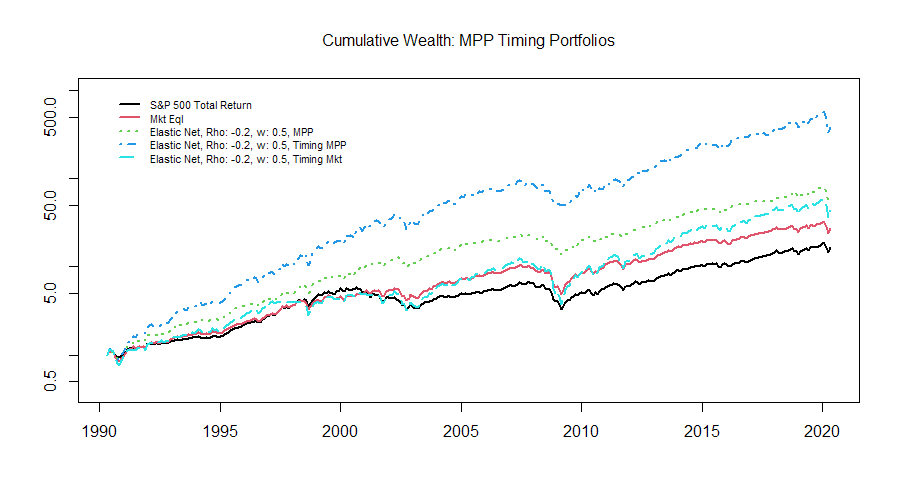}
  \caption{\textbf{Cumulative returns of Timing MPP portfolios (Elastic Net, $w_i = 0.5 $).} This figure plots the cumulative returns of the machine learning MPP portfolios from April 1990 to May 2020. The vertical axis is in log-scale.}
  \label{fig:fig2}
\end{figure}

\begin{figure}[h]
\captionsetup{labelfont=bf,font=small}
  \centering
  \includegraphics[scale=.47]{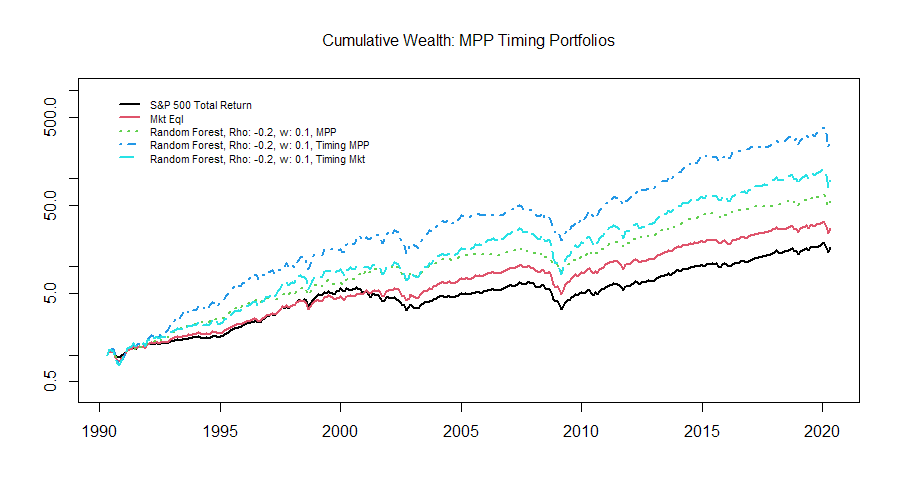}
  \caption{\textbf{Cumulative returns of Timing MPP portfolios (Random Forest, $w_i = 0.1 $).} This figure plots the cumulative returns of the machine learning MPP portfolios from April 1990 to May 2020. The vertical axis is in log-scale.}
  \label{fig:fig2}
\end{figure}

\begin{figure}[h]
\captionsetup{labelfont=bf,font=small}
  \centering
  \includegraphics[scale=.47]{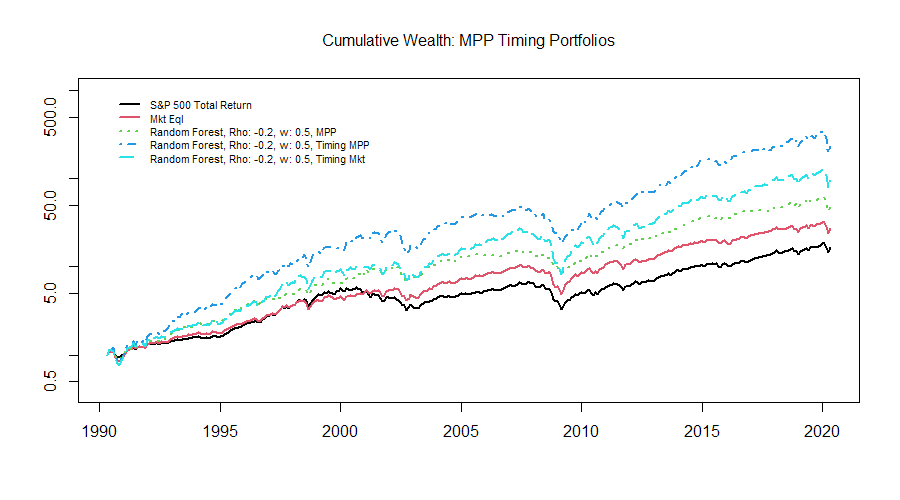}
  \caption{\textbf{Cumulative returns of Timing MPP portfolios (Random Forest, $w_i = 0.3 $).} This figure plots the cumulative returns of the machine learning MPP portfolios from April 1990 to May 2020. The vertical axis is in log-scale.}
  \label{fig:fig2}
\end{figure}

\begin{figure}[h]
\captionsetup{labelfont=bf,font=small}
  \centering
  \includegraphics[scale=.47]{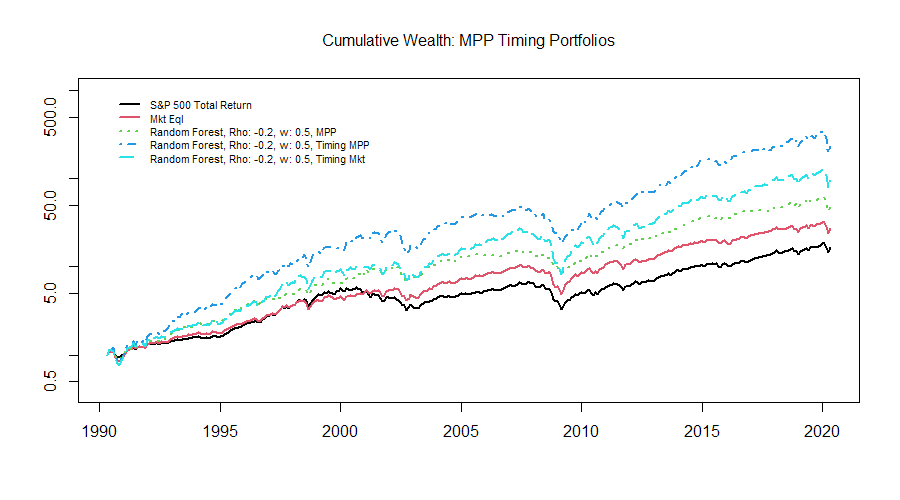}
  \caption{\textbf{Cumulative returns of Timing MPP portfolios (Random Forest, $w_i = 0.5 $).} This figure plots the cumulative returns of the machine learning MPP portfolios from April 1990 to May 2020. The vertical axis is in log-scale.}
  \label{fig:fig2}
\end{figure}

\begin{table}[!htbp] 
\captionsetup{labelfont=bf,font=normalsize}
\caption{\textbf{Sharpe Ratios}}
\justifying{\small{\noindent  ($w_i = 0.1$). In this table are the out-of-sample annual returns, standard deviations, Sharpe ratios, and turnover for the holdout period from April 1990 to May 2020. Mkt denotes the buy-and-hold.}}

\medskip

\centering
\label{tab:3}
\resizebox{\textwidth}{!}{
\begin{tabular}{lllll}
        ~ & Annual Return (\%) & Standard Deviation (\%) & Sharpe & Turnover (\%) \\ \hline
      S\&P 500 & 10.46 & 14.6 & 0.54 & - \\ 
        Mkt Eql & 12.33 & 15.95 & 0.61 & - \\ 
        Mkt Tim & 15.36 & 15.95 & 0.61 & - \\ 
        OLS MPP & 14.16 & 14.61 & 0.79 & 50.25 \\ 
        OLS Min Err & 10.26 & 12.66 & 0.6 & 48.82 \\ 
        OLS Min Err/Ret & 9.9 & 14.06 & 0.52 & 93.57 \\ 
        OLS 1st Dec Eql & 20.77 & 22.5 & 0.81 & 76.54 \\ 
        OLS 10th Dec Eql & 5.45 & 17.34 & 0.16 & 86.5 \\ 
        OLS 1st Dec MPP & 21.42 & 22.56 & 0.83 & 98.64 \\ 
        OLS 10th Dec MPP & 4.45 & 18.14 & 0.1 & 132.7 \\ 
        OLS 1st Dec Err & 17.03 & 17.44 & 0.83 & 101.29 \\ 
        OLS 10th Dec Err & 5.24 & 14.53 & 0.18 & 115.22 \\ 
        OLS 1st Dec Err/Ret & 18.62 & 18.17 & 0.88 & 96.38 \\ 
        OLS 10th Dec Err/Ret & 3.23 & 16.14 & 0.04 & 112.65 \\ 
        OLS MPP Tim & 21.56 & 21.33 & 0.89 & 2.68 \\ 
        OLS Err Tim & 13.23 & 16.17 & 0.66 & 7.97 \\ 
        OLS Err/Ret Tim & 10.43 & 14.52 & 0.54 & 26.12 \\ 
        Elastic Net MPP & 15.04 & 15.16 & 0.82 & 53.13 \\ 
        Elastic Net Min Err & 10.14 & 12.52 & 0.6 & 47.85 \\ 
        Elastic Net Min Err/Ret & 10.49 & 13.93 & 0.57 & 80.16 \\ 
        Elastic Net 1st Dec Eql & 21.71 & 23.27 & 0.82 & 71.15 \\ 
        Elastic Net 10th Dec Eql & 4.62 & 16.15 & 0.12 & 70.54 \\ 
        Elastic Net 1st Dec MPP & 23.75 & 23 & 0.92 & 94.35 \\ 
        Elastic Net 10th Dec MPP & 4.45 & 17.49 & 0.11 & 120.46 \\ 
        Elastic Net 1st Dec Err & 20.23 & 18.33 & 0.96 & 96.91 \\ 
        Elastic Net 10th Dec Err & 4.48 & 13.36 & 0.14 & 94.45 \\ 
        Elastic Net 1st Dec Err/Ret & 21.6 & 20.05 & 0.95 & 94.1 \\ 
        Elastic Net 10th Dec Err/Ret & 3.59 & 15.38 & 0.06 & 100.28 \\ 
        Elastic Net MPP Tim & 22.73 & 22.22 & 0.91 & 2.75 \\ 
        Elastic Net Err Tim & 12.36 & 16.01 & 0.61 & 7.46 \\ 
        Elastic Net Err/Ret Tim & 11.99 & 16.88 & 0.56 & 13.84 \\ 
        Random Forest MPP & 14.67 & 15.7 & 0.77 & 59.58 \\ 
        Random Forest Min Err & 9.85 & 12.15 & 0.6 & 47.16 \\ 
        Random Forest Min Err/Ret & 9.54 & 14.16 & 0.49 & 67.04 \\ 
        Random Forest 1st Dec Eql & 18.92 & 26.61 & 0.61 & 50.99 \\ 
        Random Forest 10th Dec Eql & 8.02 & 15.85 & 0.34 & 64.26 \\ 
        Random Forest 1st Dec MPP & 19.52 & 27.02 & 0.63 & 86.4 \\ 
        Random Forest 10th Dec MPP & 8.28 & 15.87 & 0.36 & 105.49 \\ 
        Random Forest 1st Dec Err & 18.56 & 21.11 & 0.76 & 76.58 \\ 
        Random Forest 10th Dec Err & 6.94 & 12.81 & 0.34 & 89.44 \\ 
        Random Forest 1st Dec Err/Ret & 18.43 & 23.97 & 0.66 & 76.44 \\ 
        Random Forest 10th Dec Err/Ret & 8.22 & 15.38 & 0.36 & 77.13 \\ 
        Random Forest MPP Tim & 21.54 & 23.05 & 0.82 & 2.29 \\ 
        Random Forest Err Tim & 14.69 & 17.36 & 0.7 & 4.54 \\ 
        Random Forest Err/Ret Tim & 12.57 & 19.72 & 0.5 & 4.33 \\ 
    \end{tabular}
}
\end{table}

\begin{table}[!htbp] 
\captionsetup{labelfont=bf,font=normalsize}
\caption{\textbf{Sharpe Ratios}}
\justifying{\small{\noindent  ($w_i = 0.5$). In this table are the out-of-sample annual returns, standard deviations, Sharpe ratios, and turnover for the holdout period from April 1990 to May 2020. Mkt denotes the buy-and-hold.}}

\medskip

\centering
\label{tab:3}
\resizebox{\textwidth}{!}{
\begin{tabular}{lllll}
        ~ & Annual Return (\%) & Standard Deviation (\%) & Sharpe & Turnover (\%) \\ \hline
        S\&P 500 & 10.46 & 14.6 & 0.54 & - \\ 
        Mkt Eql & 12.33 & 15.95 & 0.61 & - \\ 
        Mkt Tim & 15.36 & 15.95 & 0.61 & - \\ 
        OLS MPP & 13.81 & 14.66 & 0.76 & 53.21 \\ 
        OLS Min Err & 10.47 & 12.84 & 0.61 & 53.55 \\ 
        OLS Min Err/Ret & 9.06 & 14.48 & 0.45 & 98.88 \\ 
        OLS 1st Dec Eql & 20.77 & 22.5 & 0.81 & 76.54 \\ 
        OLS 10th Dec Eql & 5.45 & 17.34 & 0.16 & 86.5 \\ 
        OLS 1st Dec MPP & 21 & 22.64 & 0.81 & 103.16 \\ 
        OLS 10th Dec MPP & 4.12 & 18.25 & 0.08 & 138.27 \\ 
        OLS 1st Dec Err & 14.32 & 17.08 & 0.69 & 113.73 \\ 
        OLS 10th Dec Err & 6.12 & 15.17 & 0.23 & 129.91 \\ 
        OLS 1st Dec Err/Ret & 16.21 & 18.62 & 0.73 & 105.62 \\ 
        OLS 10th Dec Err/Ret & 3.45 & 16.97 & 0.05 & 119.99 \\ 
        OLS MPP Tim & 20.88 & 21.36 & 0.86 & 4.2 \\ 
        OLS Err Tim & 13.63 & 16.12 & 0.68 & 8.85 \\ 
        OLS Err/Ret Tim & 9.08 & 14.83 & 0.44 & 25.52 \\ 
        Elastic Net MPP & 15.06 & 15.23 & 0.82 & 55.7 \\ 
        Elastic Net Min Err & 10.41 & 12.85 & 0.61 & 54.11 \\ 
        Elastic Net Min Err/Ret & 10.78 & 14.92 & 0.55 & 88.59 \\ 
        Elastic Net 1st Dec Eql & 21.71 & 23.27 & 0.82 & 71.15 \\ 
        Elastic Net 10th Dec Eql & 4.62 & 16.15 & 0.12 & 70.54 \\ 
        Elastic Net 1st Dec MPP & 23.75 & 22.98 & 0.92 & 98.2 \\ 
        Elastic Net 10th Dec MPP & 4.12 & 17.49 & 0.09 & 125.42 \\ 
        Elastic Net 1st Dec Err & 16.82 & 17.73 & 0.8 & 108.6 \\ 
        Elastic Net 10th Dec Err & 5.32 & 13.87 & 0.19 & 108.86 \\ 
        Elastic Net 1st Dec Err/Ret & 22.02 & 20.37 & 0.95 & 106.24 \\ 
        Elastic Net 10th Dec Err/Ret & 4.74 & 15.32 & 0.14 & 112.51 \\ 
        Elastic Net MPP Tim & 22.44 & 22.13 & 0.9 & 2.96 \\ 
        Elastic Net Err Tim & 12.37 & 15.96 & 0.61 & 8.06 \\ 
        Elastic Net Err/Ret Tim & 13.36 & 17.22 & 0.62 & 16.69 \\ 
        Random Forest MPP & 14.28 & 15.87 & 0.74 & 61.4 \\ 
        Random Forest Min Err & 9.79 & 12.5 & 0.57 & 54.43 \\ 
        Random Forest Min Err/Ret & 9.54 & 15.16 & 0.46 & 72.88 \\ 
        Random Forest 1st Dec Eql & 18.92 & 26.61 & 0.61 & 50.99 \\ 
        Random Forest 10th Dec Eql & 8.02 & 15.85 & 0.34 & 64.26 \\ 
        Random Forest 1st Dec MPP & 19.31 & 27.22 & 0.61 & 89.96 \\ 
        Random Forest 10th Dec MPP & 8.35 & 15.82 & 0.36 & 107.96 \\ 
        Random Forest 1st Dec Err & 16.43 & 20.7 & 0.67 & 91.69 \\ 
        Random Forest 10th Dec Err & 6.13 & 13.39 & 0.26 & 101.24 \\ 
        Random Forest 1st Dec Err/Ret & 16.32 & 23.84 & 0.57 & 84.07 \\ 
        Random Forest 10th Dec Err/Ret & 7.46 & 15.75 & 0.31 & 80.66 \\ 
        Random Forest MPP Tim & 21.05 & 23.25 & 0.79 & 2.55 \\ 
        Random Forest Err Tim & 14.01 & 17.18 & 0.66 & 4.65 \\ 
        Random Forest Err/Ret Tim & 12.9 & 20.37 & 0.5 & 5.23 \\ 
    \end{tabular}
}
\end{table}

\end{document}